\documentclass[fleqn,twoside]{article}
\usepackage{espcrc2}

\usepackage{graphicx}
\usepackage[figuresright]{rotating}

\newcommand{\AmS}{{\protect\the\textfont2
  A\kern-.1667em\lower.5ex\hbox{M}\kern-.125emS}}

\title{The Nature of Dark Matter in Elliptical Galaxies: {\it Chandra}
Observations of NGC 4636}

\author{M. Loewenstein\address[GSFC]{Laboratory for High Energy
Astrophysics, NASA/Goddard Space Flight Center, \\ Code 662, Greenbelt,
MD 20771, USA}\thanks{also with University of Maryland Department of
Astronomy}, and R. Mushotzky\addressmark}
       
\begin{document}

\begin{abstract}
We determine the total enclosed mass profile from 0.7 to 35 kpc in the
elliptical galaxy NGC 4636 based on the hot interstellar medium
temperature profile measured using the {\it Chandra} X-ray
Observatory, and other X-ray and optical data. The total mass
increases as $r^{1.2}$ to a good approximation over this range in
radii, attaining a total of $\sim 1.5\times 10^{12}$ M$_{\odot}$
(corresponding to $M_{\rm tot}/L_V=40$) at 35 kpc. We find that at
least half, and as much as 80\%, of the mass within the optical
half-light radius is non-luminous, implying an exceptionally low
baryon fraction in NGC 4636. The large inferred dark matter
concentration and central dark matter density, consistent with the
upper end of the range expected for standard cold dark matter halos,
imply that mechanisms proposed to explain low dark matter densities in
less massive galaxies are not effective in elliptical galaxies.
\vspace{1pc}
\end{abstract}

\maketitle

\section{Context}

According to recent estimates, 80--90\% of the matter in the universe
is non-baryonic. With the presence of extended dark matter halos in
galaxies of all morphological types now well-established, attention is
focusing on comparing the detailed mass distribution with theoretical
predictions of galactic dark halo structure. The shape of the dark
matter distribution is determined by the initial density perturbation
spectrum, the coupled dynamical evolution of baryonic and non-baryonic
constituents, and the nature of the dark matter itself. Therefore, its
measurement represents a powerful diagnostic of fundamental
astrophysical processes and parameters.

The standard cold dark matter (CDM) model is highly successful in
explaining the distribution of mass in the universe on scales ranging
from galaxies on up, but is undergoing a critical re-examination --
due in large part to its confrontation with measurements of {\it
late-type} galaxy mass distributions indicating that dark matter is
less concentrated than expected. In this work \cite{LM02}, we use a
recent {\it Chandra} X-ray Observatory observation of NGC 4636 to
constrain its dark matter distribution, with the goal of testing
whether this ``concentration crisis'' applies to this ({\it
early-type}) elliptical galaxy.

\subsection{{\it Chandra} and Dark Matter Estimation}

Hot, extended distributions of X-ray emitting hot gas in hydrostatic
equilibrium are crucial tracers of the gravitational potential in
individual elliptical galaxies \cite{LW99}; gravitational lensing is
useful only in a statistical sense for a large sample of galaxies. The
physical properties of the hot gas can be accurately measured out to
large radii where dark matter dominates and optical techniques are
infeasible. The effective resolution of the mass estimate essentially
corresponds to that of the gas temperature profile (see below). The
unprecedented {\it Chandra} angular resolution enables one to do X-ray
imaging spectroscopy of the hot ISM on scales comparable to optical
studies of the stars for the first time, thus yielding the relative
distributions of luminous and dark matter.

\subsection{Modeling Approach}

Given a mass distribution, and the hot gas density distribution
obtained from the measured X-ray surface brightness distribution, we
solve the equation of hydrostatic equilibrium to derive the
corresponding gas temperature profile. The total mass model consists
of three components -- an $8\times 10^7$ M$_{\odot}$ central
supermassive black hole \cite{MF01}, stars distributed as measured
from {\it Hubble Space Telescope} and ground-based photometry, and a
dark matter density distribution parameterized by an asymptotic slope
at the origin ($\zeta$), a scale radius characterizing the transition
to an $r^{-3}$ decline (as determined by the results of numerical
simulations) at large radius, and a normalization. For a given
$\zeta$, the dark matter scale and normalization, and the stellar
mass-to-light ratio, are varied to obtain a match with the composite
temperature profile measured by {\it Chandra} and the {\it XMM-Newton}
Observatory (Figure 1).

\begin{figure}[htb]
\vspace{-20pt}
\hspace{-10pt}
\includegraphics[scale=.4]{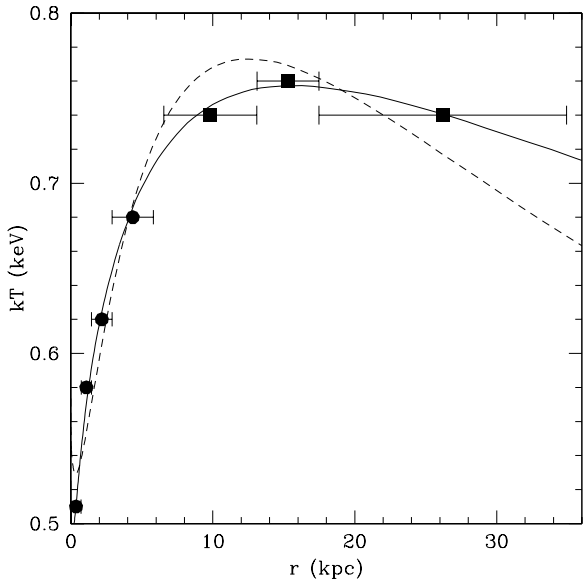}
\vspace{-40pt}
\caption{{\it Chandra} (filled circles) and {\it XMM-Newton} (filled
squares) temperatures, and profiles for overall best-fit model (which
has an $r^{-1}$ dark matter density cusp; solid curve), and best-fit
model with dark matter density core (broken curve).}
\vspace{-20pt}
\end{figure}

\section{Results}

Models with constant mass-to-light ratios and those where the dark
matter density inner slope, $\zeta$, is as steep as the singular
isothermal sphere value of $-2$ are clearly ruled out.  We derive
accurate constraints on the total mass distribution from 0.7--35 kpc
that are robust to the assumed value of $\zeta$. The total mass
increases as $r^{1.2}$ (more quickly than the stars) to a good
approximation over this range in radii, attaining a total of $\sim
1.5\times 10^{12}$ M$_{\odot}$ (corresponding to $M_{\rm tot}/L_V=40$
in solar units) at the outermost point we consider (Figure 2).

\begin{figure}[htb]
\vspace{-20pt}
\hspace{-10pt}
\includegraphics[scale=.4]{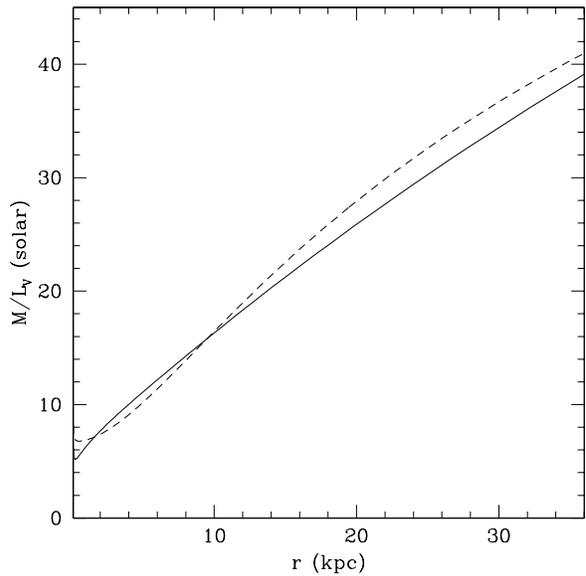}
\vspace{-40pt}
\caption{Mass-to-light ratio, in solar units, for the models
corresponding to the temperature profiles of Figure 1.}
\vspace{-20pt}
\end{figure}

There are degeneracies with respect to the relative distributions of
dark and luminous matter. We find acceptable models with dark matter
cusps (e.g., $\zeta=1, 1.5$) and with cores ($\zeta=0$). The former
imply lower stellar mass to light ratios (the allowed range: $M_{\rm
stars}/L_V=2.4-6.6$, consistent with studies of the stellar
population) and, in some cases, the predominance of dark matter all
the way into the nucleus of the galaxy. The dark matter fraction
within the half-light radius ($\sim 8$ kpc) can be as high as 0.8 and
``central'' (actually, the average over the inner 700 pc) dark matter
density as high as 4.0 M$_{\odot}$ pc$^{-3}$; the cored models define
lower limits to these quantities of 0.5 and 0.16 M$_{\odot}$
pc$^{-3}$, respectively. Even the lowest allowed central density is an
order of magnitude greater than the previously estimated values for
other types of galaxies (Figure 3) that precipitated the consideration
of alternatives to CDM.

\begin{figure}[htb]
\vspace{-20pt}
\hspace{-10pt}
\includegraphics[scale=.4]{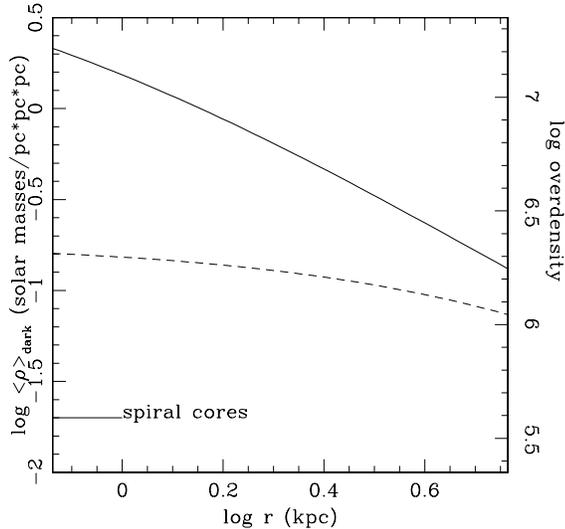}
\vspace{-40pt}
\caption{Average enclosed dark matter density distribution
corresponding to the temperature profiles of Figure 1. The typical
value estimated for spiral (and other) galaxies is indicated.}
\vspace{-20pt}
\end{figure}

\section{Implications for the Nature of Dark Matter}

Comparison of non-parametric measures of the dark matter concentration
with expectations based on standard CDM numerical simulations indicate
that dark matter in NGC 4636 is more compact than average for its mass
scale, but within the expected scatter. This is consistent with the
findings of statistical studies of the gravitational lensing
properties of elliptical galaxies, though it appears that NGC 4636 is
unusually dark matter dominated.

Many of the proposed scenarios for reducing the dark matter
concentration in late-type galaxies are at odds with our results.  It
is instructive to consider the two classes of models illustrated in
the above Figures from this perspective -- those with flat dark matter
cores, and those with cuspy dark matter cores (that provide better
fits to the X-ray data) interpreted as contracting from initially flat
cores due to baryonic infall. For the flat-core models, the high
central dark matter mass density and large cores are contrary to the
expected scaling relations for self-interacting dark matter (SIDM) and
other models where dark matter structure is driven by dark matter
particle interaction. For the cuspy-core models the central
phase-space density, conserved during adiabatic contraction, is too
high.  Alternatively, the velocity-dependent cross section can be
fine-tuned so that interactions are ineffective at the higher mass
(and velocity) scale of ellipticals; although here too, one may run
into phase space difficulties if ellipticals form from mergers of
lower mass systems with interaction-induced dark matter cores.

Scenarios where the mechanism for reducing the dark matter
concentration becomes less effective with increasing mass scale are
not ruled out, since ellipticals such as NGC 4636 represent the most
massive galaxies with the deepest potential wells. These include
models that invoke warm dark matter, or the expansion of dark matter
due to coupling with a powerful protogalactic starburst-driven
baryonic outflow. If this transitional mass is indeed on the giant
elliptical galaxy scale, cuspy dark matter distributions in galaxy
clusters are implied.

Our result that the central dark matter density in NGC 4636 is
$>8-200\times$ the typical value for less massive galaxies contradicts
the naive expectations of bottom-up hierarchical clustering where the
most massive systems form latest and reflect the relatively low
average density in the universe at that epoch. A cold dark matter
initial perturbation spectrum that is tilted or otherwise lacking in
small-scale power may help explain Ly-$\alpha$ forest data \cite{ABW},
but likely exacerbates this contradiction.

Recent work on the relation between the merging histories of galaxies
and their morphology \cite{W02} may be the key to resolving the dark
matter concentration problem. A significant dispersion -- in dark
matter concentration is predicted by numerical simulations -- at any
given mass range, but particularly on galaxy scales. This results from
the stochasticity of the merging process, and introduces the following
bias.  Galaxies of low central dark matter density represent
relatively recently formed systems, and/or particularly fragile
galaxies that experienced relatively tranquil assembly histories --
these may correspond to disk galaxies where low dark matter central
densities are indeed measured. Conversely, giant elliptical galaxies
such as, or perhaps particularly, NGC 4636 may form at relatively high
redshifts and undergo exceptionally prominent merger histories.

\section{Concluding Remarks}

Possible observational inconsistencies of the (non-interacting) CDM
paradigm resulted in a renewed scrutiny and a proliferation of
alternative models. Careful analysis of the observational situation in
light of more detailed, and more deeply understood, theoretical
models, as well as the shortcomings of the proposed alternatives seem
to indicate that CDM is withstanding these recent challenges to remain
the most viable model of structure in the universe (see J. Primack's
contribution to this volume). Our results on the high concentration of
dark matter in NGC 4636 evidently strengthens the case for CDM, while
introducing other puzzles -- such as why NGC 4636 seems to be
exceptionally dark matter dominated. Although one must exercise
caution in generalizing from our results, the most stringent
constraints on alternative models for dark halo structure, that
presumably universally apply, emerge from studying such a, possibly
extreme, system. Results, in progress, of similar investigations for
other galaxies (e.g., NGC 1399, NGC 4472) are anticipated with great
interest.

\end{document}